\documentclass{article}
\usepackage{PRIMEarxiv}
\usepackage[utf8]{inputenc} % allow utf-8 input
\usepackage[T1]{fontenc} % use 8-bit T1 fonts
\usepackage{hyperref} % hyperlinks
\usepackage{url} % simple URL typesetting
\usepackage{booktabs} % professional-quality tables
\usepackage{amsfonts} % blackboard math symbols
\usepackage{nicefrac} % compact symbols for 1/2, etc.
\usepackage{microtype} % microtypography
\usepackage{lipsum}
\usepackage{fancyhdr} % header
\usepackage{graphicx} % graphics
\usepackage{xcolor} % For colors
\usepackage[normalem]{ulem} % For strikethrough (\sout)
\graphicspath{{media/}} % organize your images and other figures under media/ folder
\usepackage[numbers]{natbib} % or [authoryear]
\usepackage{caption} % For Figure caption
\usepackage{float} % For [H] placement
\usepackage{longtable} % For affiliation table

\usepackage[compact]{titlesec}
\titlespacing*{\section}{0pt}{*2.0}{*0.8}
\titlespacing*{\subsection}{0pt}{*1.6}{*0.6}
\titlespacing*{\subsubsection}{0pt}{*1.2}{*0.4}
\usepackage{etoolbox}
\apptocmd{\thebibliography}{\setlength{\itemsep}{2pt}\setlength{\parskip}{0pt}}{}{}
\title{How malicious AI swarms can threaten democracy
\\[0.5ex]
\large\textit{The fusion of agentic AI and LLMs marks a new frontier in information warfare.}
}

\author{
\parbox{\linewidth}{
\centering
\normalsize
\textbf{Daniel Thilo Schroeder}\textsuperscript{1\dag*}, %
\textbf{Meeyoung Cha}\textsuperscript{2}, %
\textbf{Andrea Baronchelli}\textsuperscript{3}, %
\textbf{Nick Bostrom}\textsuperscript{4}, %
\textbf{Nicholas A.~Christakis}\textsuperscript{5}, %
\textbf{David Garcia}\textsuperscript{6}, %
\textbf{Amit Goldenberg}\textsuperscript{7}, %
\textbf{Yara Kyrychenko}\textsuperscript{8}, %
\textbf{Kevin Leyton-Brown}\textsuperscript{9}, %
\textbf{Nina Lutz}\textsuperscript{10}, %
\textbf{Gary Marcus}\textsuperscript{11}, %
\textbf{Filippo Menczer}\textsuperscript{12}, %
\textbf{Gordon Pennycook}\textsuperscript{13}, %
\textbf{David G.~Rand}\textsuperscript{14}, %
\textbf{Maria Ressa}\textsuperscript{15}, %
\textbf{Frank Schweitzer}\textsuperscript{16}, %
\textbf{Dawn Song}\textsuperscript{17}, %
\textbf{Christopher Summerfield}\textsuperscript{18}, %
\textbf{Audrey Tang}\textsuperscript{19}, %
\textbf{Jay J.~Van~Bavel}\textsuperscript{11,20}, %
\textbf{Sander van~der~Linden}\textsuperscript{8}, %
\textbf{\& Jonas R.~Kunst}\textsuperscript{21\dag} %
\\[2ex]
\scriptsize
\normalfont
\textsuperscript{1}Department of Sustainable Communication Technologies, SINTEF Digital; Oslo, Norway; %
\textsuperscript{2}Max Planck Institute for Security and Privacy; Bochum, Germany; %
\textsuperscript{3}Department of Mathematics, City St George’s University of London; London, England, UK; %
\textsuperscript{4}Macrostrategy Research Initiative; London, England, UK; %
\textsuperscript{5}Human Nature Lab, Yale University; New Haven, Connecticut, USA; %
\textsuperscript{6}Department of Politics and Public Administration, University of Konstanz; Konstanz, Germany; %
\textsuperscript{7}Harvard Business School, Harvard University; Boston, Massachusetts, United States; %
\textsuperscript{8}Department of Psychology, University of Cambridge; Cambridge, England, United Kingdom; %
\textsuperscript{9}Department of Computer Science, University of British Columbia; Vancouver, British Columbia, Canada; %
\textsuperscript{10}Department of Human Centered Design \& Engineering, University of Washington; Seattle, Washington, USA; %
\textsuperscript{11}Department of Psychology, New York University; New York City, New York, USA; %
\textsuperscript{12}Observatory on Social Media and Luddy School of Informatics, Computing, and Engineering, Indiana University; Bloomington, Indiana, USA; %
\textsuperscript{13}Department of Psychology, Cornell University; Ithaca, New York, USA; %
\textsuperscript{14}Department of Information Science and Marketing and Management Communication, Cornell University; Ithaca, New York, USA; %
\textsuperscript{15}School of International and Public Affairs, Columbia University, New York, NY USA \& Rappler.com; Philippines; %
\textsuperscript{16}Department of Management, Technology, and Economics, ETH Zürich; Zurich, Switzerland; %
\textsuperscript{17}Department of Electrical Engineering and Computer Science, University of California Berkeley; Berkeley, California, USA; %
\textsuperscript{18}Department of Experimental Psychology, University of Oxford; Oxford, England, United Kingdom; %
\textsuperscript{19}Cyber Ambassador Taiwan; Taipei, Taiwan; %
\textsuperscript{20}Department of Strategy and Management, Norwegian School of Economics; Bergen, Norway; %
\textsuperscript{21}Department of Communication and Culture, BI Norwegian Business School; Oslo, Norway. \\
\textbf{Email: daniel.t.schroeder@sintef.no}
}
}

\makeatletter

\renewcommand{\@toptitlebar}{%
  \vskip 0.25in
}
\renewcommand{\@bottomtitlebar}{%
  \vskip 0.1in
}

\renewcommand{\@maketitle}{%
  \vbox{%
    \hsize\textwidth
    \linewidth\hsize
    \vskip 0.35in                 % top margin before title
    \centering
    {\LARGE\bfseries \@title \par}
    \vskip 0.15in                 % space between title and subtitle
    \vskip 0.15in \@minus 0.05in  % gap title → authors (kept tight as before)
    {\@author}
    \vskip 0.60in                 % ← increased from 0.2in → more space before main text
  }%
}

\makeatother

\begin{document}

%%%%%%%%%%%%%%%%%%%%%%%%%%%%%%%%%%%%%%%%%%%%%%%%%%%%%%%%%%%%%%%%%%%%
\vspace*{0.6\baselineskip}   % small space from top margin – adjust if needed
\noindent
\footnotesize
\textcolor{red}{%
This is the author's version of the work. It is posted here by permission of the AAAS for personal use, not for redistribution. \\
The definitive version was published in Science on January 22, 2026, DOI: 10.1126/science.adz1697.
}\par
\vspace{1.0\baselineskip}   % vertical gap between notice and title – tune this value
\maketitle
\vspace*{-2.0\baselineskip}
%%%%%%%%%%%%%%%%%%%%%%%%%%%%%%%%%%%%%%%%%%%%%%%%%%%%%%%%%%%%%%%%%%%%

%%%%%%%%%%%%%%%%%%%%%%%%%%%%%%%%%%%%%%%%%%%%%%%%%%%%%%%%%%%%%%%%%%%%
\maketitle
\vspace*{-2.0\baselineskip}
%%%%%%%%%%%%%%%%%%%%%%%%%%%%%%%%%%%%%%%%%%%%%%%%%%%%%%%%%%%%%%%%%%%%
%%%%%%%%%%%%%%%%%%%%%%%%%%%%%%%%%%%%%%%%%%%%%%%%%%%%%%%%%%%%%%%%%%%%
Advances in AI offer the prospect of manipulating beliefs and behaviors on a population-wide level~\cite{bengio2024}. Large language models (LLMs) and autonomous agents~\cite{wang2024} now let influence campaigns reach unprecedented scale and precision. Generative tools can expand propaganda output without sacrificing credibility~\cite{wack2025} and inexpensively create falsehoods that are rated as more human-like than those written by humans~\cite{williams2025,wack2025}. Techniques meant to refine AI reasoning, such as chain-of-thought prompting, can just as effectively be used to generate more convincing falsehoods. Enabled by these capabilities, a disruptive threat is emerging: swarms of collaborative, malicious AI agents. Fusing LLM reasoning with multi-agent architectures~\cite{wang2024}, these systems are capable of coordinating autonomously, infiltrating communities, and fabricating consensus efficiently. By adaptively mimicking human social dynamics, they threaten democracy. Because the resulting harms stem from design, commercial incentives, and governance, we prioritize interventions at multiple leverage points, focusing on pragmatic mechanisms over voluntary compliance.

This risk compounds long-standing vulnerabilities in democratic information ecosystems, already weakened by erosion of rational-critical discourse and a lack of shared reality among citizens. AI swarms are a potent accelerant in this trajectory, though their ultimate impact is not predetermined. Their effects will be shaped by platform design, market incentives, media institutions, and political actors. Here, we distinguish documented trends from projections, indicate where uncertainty remains, and note countervailing dynamics, such as growing public skepticism toward unverified content and a renewed interest in institutional demand for accountable journalism (see supplementary materials). 

AI swarms continue a long history of communication technologies reshaping political power. The advent of the printing press enabled a ``public sphere'' and the mass circulation of ideas that challenged state authority. The broadcast era centralized influence in a one-to-many communication model. Here, the public sphere shifted from a site of participation to one of mass media consumption, often exploited by politicians and their parties for national cohesion and mass persuasion. The digital era then fragmented this landscape. By lowering entry barriers, social media platforms enabled many-to-many communication and simultaneously a polarized environment for modern information operations. In this context, online manipulation accelerated, driven increasingly by domestic political elites and parties (now understood to be major drivers of disinformation) alongside foreign state actors. They have targeted events like Brexit and elections in the U.S., Brazil, and the Philippines. Our backdrop is thus not an idealized public sphere, but one strained by decades of technological disruption and democratic backsliding. This has manifested as a sharp decline in public trust in core institutions (including the media, science, and government), thereby corroding the very foundations of evidence-based discourse upon which democratic deliberation depends.

A prime pre-generative AI example of influence operations is the state-backed, human-driven botnet. During the Russian Internet Research Agency's (IRA) 2016 Twitter operation, only one percent of users saw 70\% of its content, with no detectable effects on opinions or turnout~\cite{eady2023}. We do not claim that the IRA ``failed'' entirely due to technical shortcomings; its objectives also included sowing epistemic uncertainty and distrust. Nevertheless, this example highlights the cost, cadence, and iteration limits inherent to human-operated systems that new developments in AI can help overcome.

This leap, from human- to AI-driven influence operations, is underway. LLMs generate persuasive, tailored text at scale and have shifted deep-seated beliefs in laboratory settings~\cite{costello2024}. Open-source releases further lower access barriers. Consequently, AI-supported election interference is no longer hypothetical. Taiwan's, India's, Indonesia's, and the U.S.' 2024 campaigns saw deepfakes, and fabricated news outlets now influence debates. Absent guardrails, LLM-driven swarms can transform sporadic mis- and disinformation into persistent, adaptive manipulation of democratic discourse.

%%%%%%%%%%%%%%%%%%%%%%%%%%%%%%%%%%%%%%%%%%%%%%%%%%%%%%%%%%%%%%%%%%%%
\section*{Swarm Capabilities}
%%%%%%%%%%%%%%%%%%%%%%%%%%%%%%%%%%%%%%%%%%%%%%%%%%%%%%%%%%%%%%%%%%%%
A malicious AI swarm is a set of AI-controlled agents that (i) maintains persistent identities and memory; (ii) coordinates toward shared objectives while varying tone and content; (iii) adapts in real time to engagement, platform cues, and human responses; (iv) operates with minimal human oversight; and (v) can deploy across platforms. Classic coordinated inauthentic behavior amplifies the spread of information by inflating content frequency and engagement to trigger algorithmic visibility through repetition, manual scheduling, and rigid scripts. Swarms differ by fusing scale, heterogeneity, and real-time adaptation: they can generate organic-looking, context-aware content, sustain coherent narratives across agents, and evolve with feedback. This synthesis, enabled by model-driven generation, memory, and planning, could achieve effects that conventional, human-intensive operations cannot match in speed or cost.

Recent breakthroughs in multi-agent systems have fused LLM reasoning with agentic memory, planning, and communication~\cite{park2023}. Five advances now matter for influence operations.

First is the shift from central command to fluid, real-time coordination. A single adversary could operate thousands of AI personas, scheduling content and updating narrative frames across fleets. Local adaptation plus periodic synchronization with a central node blurs the line between command-and-control and emergent ``hive'' behavior. If these agent swarms evolve into loosely governed ``societies,'' with internal norm formation and division of labor, the challenge shifts from tracing commands to understanding emergent group cognition~\cite{flint2025}. These ``societies'' may undergo spontaneous or adversarially-induced norm shifts, abandoning engineered constraints for new behavioral patterns via tipping-point effects~\cite{flint2025}.

Second, agents can employ systems that map social network structures at scale and infiltrate vulnerable communities with tailored appeals, winning followers~\cite{truong2024}. They can identify key communities and beliefs, and track trending topics. This process can be decentralized with global, network-wide efficacy~\cite{shirado2017}. Equipped with such capabilities, swarms can position for maximum impact and tailor messages to the beliefs and cultural cues of each community, enabling more precise targeting than previous botnets.

Third, human-level mimicry helps swarms evade detectors that once caught simpler ``copy-paste'' bots. Detection of coordinated inauthentic behavior generally relies on activity patterns being suspiciously similar across accounts and, thus, statistically unlikely to be independent~\cite{pacheco2021}. Photorealistic avatars, context-appropriate slang, and heterogeneous posting rhythms can circumvent the synchrony older detectors flag.

Fourth, swarms may become increasingly self-optimizing, harvesting real-time engagement data, recommender cues, or user feedback in plain language. With sufficient signals, they may run millions of micro-A/B tests, propagate the winning variants at machine speed, and iterate far faster than humans.

Finally, an around-the-clock presence turns influence into a long-term, low-friction infrastructure. Unlike transient operations, agent swarms can persist, embedding themselves within communities over long timescales and gradually shifting discourse. This persistent influence can drive deeper cultural changes beyond norm shifts, subtly altering a community's language, symbols, and identity~\cite{brinkmann2023}. It amplifies other mechanisms described above. In cognitive warfare, AI's relentless operational endurance becomes a weapon against limited human efforts.

%%%%%%%%%%%%%%%%%%%%%%%%%%%%%%%%%%%%%%%%%%%%%%%%%%%%%%%%%%%%%%%%%%%%
\section*{Pathways of Harms to Democracy}
%%%%%%%%%%%%%%%%%%%%%%%%%%%%%%%%%%%%%%%%%%%%%%%%%%%%%%%%%%%%%%%%%%%%
Emerging capabilities of swarm-driven influence campaigns threaten democracy by shaping public opinion, which leads to cascading harms. These pathways are conditional claims that may materialize, especially where recommenders, ad markets, and moderation practices reward coordinated messaging with weak provenance, and where business models privilege engagement over authenticity (i.e., the currently dominant model of most social media platforms). Emerging counter-trends such as migration to smaller communities and increased reliance on verified outlets may mitigate some harms.

In today's fragmented information environment, ideological echo chambers offer fertile ground for manipulation. AI swarms are uniquely equipped to exploit this by engineering a synthetic consensus that appears to bridge these divides. They may seed narratives across disparate niches, creating an illusion of majority agreement. They can also boost this illusion by liking posts, making narratives appear widely supported. Citizens then update opinions based on peer norms, more so than evidence. A chorus of seemingly independent voices creates a mirage of bipartisan grassroot consensus with enhanced speed and persuasiveness. The result is deeply embedded manipulation that lets operators nudge public discourse almost invisibly over time.

This chorus erodes the independence essential to collective intelligence and democracy, already weakened by pervasive social influence operations on contemporary platforms. Beyond social norms, this directly undermines human cognitive information processing. The ``wisdom of crowds,'' where aggregated judgments outperform experts, depends critically on independence between judgments. While rudimentary botnets already replicate messages to simulate consensus, swarms of AI agents can do so with far greater sophistication, adaptivity, and contextual awareness. Citizens may then overestimate the informational value of this artificial consensus and may further magnify it by sharing the information themselves. Coordinated outputs can erode independence and diversity of inputs, particularly when platform features amplify social proof and herd signals; where governance or platform design reduces these incentives, effects may attenuate.

Collaborating agents can tailor misleading information to each sub-community's linguistic, cultural, and emotional markers, weaving segmented realities. These engineered realities can be designed to keep groups apart, making cross-cleavage consensus less feasible. Once initiated, such streams can spread via social contagion, with the effect of agents potentially cascading beyond direct connections.

By flooding the web with fabricated chatter, swarms can contaminate training data. This long-term ``LLM Grooming'' strategy allows adversaries to poison the epistemic substrate of AI. This threat is not theoretical: analysis of pro-Kremlin influence operations like the ``Pravda'' network suggests such tactics are already in use. These networks appear purpose-built for machine consumption. Duplication of articles across hundreds of domains, poor user interfaces, and low human traffic indicate their primary audience is web crawlers feeding LLMs. Operators deploy faux publics that flood the web. LLMs then ingest this chatter; at the next retraining cycle, fabricated narratives calcify in model weights~\cite{bowen2024}. Thus, AI swarms can rig the epistemic substrate on which future deliberation and future AI tools will rely, undermining the informed public deliberation on which democracy depends.

Separate from fragmentation, swarms can cheaply unleash coordinated synthetic harassment that relentlessly targets politicians, dissidents, academics, whistleblowers, journalists, and their networks with overwhelming, tailored abuse. Unlike conventional trolling, these swarms appear spontaneous while actually orchestrated by thousands of AI personas adapting to target responses. By the time monitoring teams distinguish AI campaigns from organic criticism, targets may have withdrawn from public life, delivering substantial victories for campaign operators while systematically excluding critical voices from democratic discourse.

As trust, already declining in many contexts, collapses, fear, uncertainty, and doubt (FUD) can drive users into gated channels and silence. When citizens realize that vast portions of online speech may be AI-generated, trust in platforms and users declines further. This shift is underway and has mixed implications: private groups can improve context, norms, and safety. Yet, they may reduce cross-cutting exposure, interfere with democratic speech, and transfer moderation from public to private actors -- a trade-off that avoids centralized state control but raises concerns about opacity and uneven enforcement.

Some threat actors may even welcome their synthetic interventions being exposed, reasoning that exposing manipulation can sow as much confusion as successful deception. Compounding this, users may be misidentified as bots, weaponizing false accusations to discredit individuals and intensify FUD. This ``epistemic vertigo'' may mesh with low-cost LLM spam that overwhelms social media feeds, making human conversation harder to find. Together, FUD and content saturation could drive disengagement, shrinking the shared public sphere on which democracy relies. This trajectory is constrained by a critical boundary condition: given that mass user disengagement threatens platforms' business models that depend on engagement, they will be incentivized to intervene. Their objective, however, would likely not be elimination but calibration to balance maximum engagement with stability.

Algorithmic over-compensation can then elevate celebrity and elite voices while sidelining ordinary citizens. When feeds flood with AI-authored posts, both ranking algorithms and users may retreat to trust proxies, such as the number of followers, official verification badges, and pre-existing traditional fame. Attention may concentrate around influencers, political elites, celebrities, and major brands, while ordinary participants fade. The public sphere contracts from many-to-many dialogue back to a few-to-many broadcast, eroding democratic pluralism and encouraging cynicism or migration to closed groups. Simultaneously, renewed public trust in professional journalism and greater reliance on accountable, verified outlets can improve attribution and reduce noise. Thus, whether this concentration of attention and influence represents a democratic loss or resilience gain may depend on access, pluralism, and transparency within those institutions.

Swarms may tip norms into action or dampen conformity, accelerating anti-democratic action~\cite{centola2018}. Rather than occupying central or influential positions, these agents could operate on the periphery of social networks, where early mobilization often begins~\cite{barbera2015}. Similar strategies can be weaponized for micro-targeted voter suppression or mobilization. Reinforcement-learning agents could run thousands of experiments per hour, iteratively adjusting content while mining engagement and responses to infer voting intent and tactic success.

Taken to extremes, coordinated doubt may corrode institutional legitimacy and invite ``emergency'' rule. By coordinating subtle, growing doubts about electoral commissions, courts, or statistics bureaus, swarms could corrode procedural trust. As confidence falters, ``emergency'' measures (e.g., postponing elections, rejecting certified results) may become palatable, especially if deepfake endorsements from fabricated civic leaders amplify the call.

%%%%%%%%%%%%%%%%%%%%%%%%%%%%%%%%%%%%%%%%%%%%%%%%%%%%%%%%%%%%%%%%%%%%
\section*{Governance Measures and Technical Defenses}
%%%%%%%%%%%%%%%%%%%%%%%%%%%%%%%%%%%%%%%%%%%%%%%%%%%%%%%%%%%%%%%%%%%%
The emergence of AI swarms marks a critical juncture. Causality runs both ways: swarms endanger democratic norms, and governance quality shapes how potent or containable swarms become. While the escalating harms may lead some to advocate for abandoning these platforms altogether, their integration into modern social, political, and economic life makes widespread disengagement unlikely. The challenge we focus on here, therefore, is not how to dismantle platforms, but how to fortify them against manipulation for those who will continue to rely on them. Addressing this threat requires a multi-layered approach, yet we recognize that any proposed solution faces considerable political hurdles. Domestic political elites are often among the most prolific sources of misleading or manipulative information and may be unwilling to constrain technologies they perceive as beneficial to their own campaigns and objectives. Furthermore, technology companies and their leaders may refuse to implement meaningful changes because they prioritize expansion over safety and for fear of alienating major political actors and facing partisan backlash. These political challenges are compounded by complex issues of jurisdiction and enforcement. 

Distinguishing malicious AI coordination from genuine, often bursty human grassroots coordination is a challenge. The line blurs further in grey areas where personal AI could have benign applications. For instance, tools used with clear disclosure and without impersonation might broaden civic engagement by helping users overcome barriers like language proficiency or lack of time. This raises a critical question: why can't pro-social swarms simply counter malicious ones in a symmetrical arms race? The digital attention economy often rewards content that triggers outrage, fear, and group identity (the primary tools of manipulators), making it more viral than nuanced or civil messages. Furthermore, pro-social actors are bound by ethical constraints against using the tactics (deception, impersonation, and emotional exploitation of human biases) that make malicious swarms effective. These factors might skew the emerging social dynamics in a negative direction. 

Defense is a persistent arms race between detection and evasion. Therefore, the primary goal of technical defenses is not foolproof prevention but to raise the stakes for attackers by increasing their operational complexity and resource requirements, while making discovery both more likely and more costly for them. The first line of defense should be always-on detection with public audits. Platforms and regulators could require continuous, real-time monitoring detectors that scan live traffic for statistically anomalous coordination patterns -- the imperfect fingerprints of inauthentic swarms~\cite{pacheco2021}. This focus on inauthentic behavior (i.e., provenance and coordination), rather than the semantic content of speech, would avoid the intractable role of a central arbiter of truth. By prioritizing procedural legitimacy (authentic, independent actors) over semantic truth, this framework sidesteps the deep epistemic question of who determines misinformation (note, however, that professional fact-checkers have proven to be remarkably accurate and consistent). Advanced analytics can (i) identify emergent agent clusters by surfacing camouflaged indicators of coordinated activity; and (ii) spot narrative-alignment drifts. However, attackers will inevitably adapt, for instance, by training swarms to mimic the statistical patterns of genuine grassroots mobilization, necessitating continuous evolution of defenses.

Deploying these detection systems would require mandates, audits, and transparency to prevent misuse. Relying purely on voluntary measures may be insufficient as the assumption that market forces alone will punish platforms overlooks critical market failures. Platforms often face misaligned incentives, since inauthentic accounts can inflate the engagement metrics that drive revenue, while users frequently cannot distinguish sophisticated bots from genuine activity, preventing them from effectively 'voting with their feet'. However, acknowledging market failure should not obscure the symmetric risk of government failure. Poorly designed mandates could be politically weaponized to selectively punish platforms, enforced via biased judgment calls, or implemented in ways that preemptively stifle architectural innovation (such as decentralized, protocol-based approaches). For this reason, compliance may be mandated and enforced through commercial-incentive levers, such as delisting non-compliant platforms from ad markets or app stores, thereby shifting from voluntary promises to financial consequences.

To extend protection to end users, platforms should offer optional ``AI shields.'' Shields could label posts that carry high swarm-likelihood scores, let users down-rank or hide them, and surface short provenance explanations in situ. Local scoring would preserve privacy while giving citizens agency over their information diets. Aggregated, anonymized feedback can be shared publicly, forming a distributed early-warning grid, yet this system remains vulnerable to adversarial manipulation by swarms programmed to whitelist their own propaganda and blacklist legitimate opponents through false reporting.

Simulation can stress-test detectors. Real-time monitors would be effective only when they anticipate future tactics. Because defenders lack access to the autonomous and evolving decision-making logic of AI swarms, agent-based simulation may be the only reliable window into how these systems behave. AI agents seeded into synthetic networks can replicate a platform's graph structure, content cadence, and recommender logic, yielding traces to recalibrate detectors. By repeatedly testing defenses against simulated swarms, researchers could identify the limits of their persuasive power, uncover their longer-term strategies, and reinforce protective measures.

Where manipulation slips through, calibrated defensive agents could deploy watermarked counter-narratives overtly labeled to clearly attribute them to their source. This is perhaps the most perilous countermeasure, as state-sanctioned tools for speech intervention are inherently political and risky. In the hands of a government, such tools could suppress dissent or amplify incumbents. Therefore, the deployment of defensive AI can only be considered if governed by strict, transparent, and democratically accountable frameworks. These must include independent oversight, publicly auditable criteria for what constitutes a manipulative campaign, and clear, unambiguous watermarking of all defensive content. Under such strict governance, defensive AI agents can disseminate accurate information, warn targeted communities, and promote media literacy at scale~\cite{costello2024}. Crucially, while acknowledging the asymmetric battlefield, such counter-narratives need not be merely reactive; they could also be deployed proactively to inoculate communities against emerging threats, aiming to minimize polarization and misinformation before a campaign takes hold. Counter-messaging must prioritize precision over volume; if defensive agents indiscriminately flood a platform, human voices could vanish into synthetic content, triggering the collapse we seek to avert. Thus, defensive AI should intervene only where manipulation is detected and verified.

The adaptive nature of AI swarms underscores the need for a complementary approach: strengthening provenance. Stronger provenance may reinforce the reliability of identity signals without muting speech. Policymakers may incentivize the rapid adoption of passkeys, cryptographic attestations, and federated reputation standards, backed by anti-spoofing R\&D. However, ``proof-of-human'' is no panacea: millions of people online lack identification, biometrics raise privacy risks, and verified accounts can be hijacked. Real-identity policies may deter bots, yet endanger political dissidents, activists, and whistleblowers who rely on anonymity to speak safely. Nevertheless, provenance strengthening is among the most promising ways to raise the cost of mass manipulation. Safeguards could allow verified-yet-anonymous posting, periodic re-verification to curb hijacking, and symbolic subscription fees to deter botnets. Cryptographic tools can further protect privacy while preserving accountability.

To counter the speed, scale, independence, and adaptability of AI swarms, a step toward global coordination could be a distributed `AI Influence Observatory' ecosystem -- a network of academic groups, NGOs, and multilateral institutions. Its goal would be to standardize evidence, improve situational awareness, and enable faster collective response rather than impose top-down reputational penalties. To be practical, the ecosystem should rely on narrowly defined, privacy-preserving inputs and provide vetted researcher sandboxes for independent analysis. Civil-society reporting, investigative journalism, and whistleblower channels would complement technical signals, enabling triangulation across diverse evidence streams. For severe cross-border incidents, an impartial multilateral investigatory mechanism could evaluate claims and publish verified incident reports. The observatory's verified incident reports could then serve as an impartial evidence base, enabling national or regional regulators to more effectively apply their own enforcement actions and economic sanctions.

Because regulation and voluntary compliance face considerable political resistance, and because AI swarms make sophisticated manipulation cheaper and more effective, a pragmatic approach should target underlying economic drivers. A key priority here would be to disrupt the commercial market that underpins large-scale manipulation, where private sellers offer services ranging from boosting vanity metrics to executing coordinated influence operations at remarkably low costs. Beyond detection, commercial-incentive levers can reduce profits from manipulation by domestic and foreign operators. Policies that may be helpful include adopting no-revenue policies for malicious swarm-proliferated content, discounting synthetic engagement in ranking and revenue-sharing, and publishing audited bot-traffic metrics. Safeguards must cover parties, campaigns, and officeholders, including party-linked media and contractors.

Finally, companies should be required to promptly disclose when an account is flagged for behavior indicative of coordinated inauthentic activity, ensuring transparency while allowing for processes to address potential false positives. Policymakers should encourage -- and where appropriate, incentivize -- platforms to provide meaningful, privacy-preserving access for independent researchers so research can keep pace with evolving threats. At the same time, pre-bunking campaigns can help build cognitive resilience by empowering people and systems (``model immunization'') to spot the fingerprints of AI swarms. To strengthen structural defenses, interoperable ``pro-user media'' architectures, defined by empowering design principles that prioritize user well-being and epistemic health over maximizing viral engagement, can promote healthier information flows. At the same time, governments and technology firms should prioritize AI-safety research and fund independent measurement of misuse and societal impact. Taken together, these measures offer a layered strategy: immediate transparency to restore trust, proactive education to bolster citizens, resilient infrastructures to reduce systemic vulnerabilities, and sustained investment to monitor and adapt over time.

The next few years give an opportunity to proactively manage the challenges of the next generation of AI-enabled influence operations. If platforms deploy swarm detectors, frontier labs submit models to standardized persuasion ``stress-tests,'' and governments launch an AI Influence Observatory that publishes open incident telemetry, we may be able to mitigate the most significant risks before key political future events, without freezing innovation. Doing so will require rapid iteration, data-sharing, and coordination across scientific, civil society, industry, and election security. Success depends on fostering collaborative action without hindering scientific research, while ensuring that the public sphere remains both resilient and accountable. By committing now to rigorous measurement, proportionate safeguards, and shared oversight, upcoming elections could even become a proving ground for, rather than a setback to, democratic AI governance.

%%%%%%%%%%%%%%%%%%%%%%%%%%%%%%%%%%%%%%%%%%%%%%%%%%%%%%%%%%%%%%%%%%%%
%%%%%%%%%%%%%%%%%%%%%%%%%%%%%%%%%%%%%%%%%%%%%%%%%%%%%%%%%%%%%%%%%%%%
\bibliographystyle{science}

\begin{thebibliography}{99}
\bibitem{bengio2024} Y. Bengio et al., \textit{Science} \textbf{384}, 842 (2024).
\bibitem{wang2024} L. Wang et al., Frontiers of Computer Science 18, 186345 (2024).
\bibitem{wack2025} M. Wack, C. Ehrett, D. Linvill, P. Warren, \textit{PNAS Nexus} \textbf{4}, pgaf083 (2025).
\bibitem{williams2025} A. R. Williams et al., \textit{PloS one} \textbf{20}, e0317421 (2025).
\bibitem{eady2023} G. Eady et al., \textit{Nature communications} \textbf{14}, 62 (2023).
\bibitem{costello2024} T. H. Costello, G. Pennycook, D. G. Rand, \textit{Science} \textbf{385}, eadq1814 (2024).
\bibitem{park2023} J. S. Park et al., Proc. 36th ACM Symp. User Interface Softw. Technol. (UIST ’23), 1–22 (2023).
\bibitem{flint2025} A. Flint Ashery, L. M. Aiello, A. Baronchelli, \textit{Science Advances} \textbf{11}, eadu9368 (2025).
\bibitem{truong2024} B. T. Truong, X. Lou, A. Flammini, F. Menczer, \textit{PNAS nexus} \textbf{3}, 258 (2024).
\bibitem{shirado2017} H. Shirado, N. A. Christakis, \textit{Nature} \textbf{545}, 370 (2017).
\bibitem{pacheco2021} D. Pacheco et al., Proc. Int. AAAI Conf. Web Soc. Media 15, 455–466 (2021).
\bibitem{brinkmann2023} L. Brinkmann et al., \textit{Nat. Hum. Behav.} \textbf{7}, 1855 (2023).
\bibitem{bowen2024} D. Bowen et al., Data Poisoning in LLMs: Jailbreak-Tuning and Scaling Laws. arXiv 2408.02946 [cs.CR] (2024).
\bibitem{centola2018} D. Centola, J. Becker, D. Brackbill, A. Baronchelli, \textit{Science} \textbf{360}, 1116 (2018).
\bibitem{barbera2015} P. Barberá et al., \textit{PloS one} \textbf{10}, e0143611 (2015).
\end{thebibliography}
% or \bibliographystyle{unsrtnat} / plainnat / science — depending on your preference

%%%%%%%%%%%%%%%%%%%%%%%%%%%%%%%%%%%%%%%%%%%%%%%%%%%%%%%%%%%%%%%%%%%%
%%%%%%%%%%%%%%%%%%%%%%%%%%%%%%%%%%%%%%%%%%%%%%%%%%%%%%%%%%%%%%%%%%%%

%%%%%%%%%%%%%%%%%%%%%%%%%%%%%%%%%%%%%%%%%%%%%%%%%%%%%%%%%%%%%%%%%%%%
\section*{ACKNOWLEDGMENTS}
%%%%%%%%%%%%%%%%%%%%%%%%%%%%%%%%%%%%%%%%%%%%%%%%%%%%%%%%%%%%%%%%%%%%
Authors thank J. Roozenbeek and A. Følstad for valuable input to this paper. Authors are also grateful to P. Antosz, Ö. Gürcan, I. Puga Gonzalez, and P. Tinn for fruitful discussions that contributed to the development of this work. AI tools (Grammarly, OpenAI o3, Claude) were used to improve the language of the manuscript. K. L.-B. is a consultant to AI21 Labs, an affiliate of Auctionomics, Inc., and an advisor to OneChronos. M. R. is the CEO and co-founder of Rappler and the founder of The Nerve, a data forensics and research consultancy. D.T.S and J.R.K. contributed equally to this work. The authors are listed in alphabetical order by last name, starting from the third and ending with the second-to-last.

\end{document}